\documentclass[10pt, final,twocolumn] {IEEEtran}

\usepackage{epsfig,cite}
\usepackage{setspace}
\usepackage{subfigure}
\usepackage{stfloats}

\usepackage{graphicx}
\usepackage{longtable}
\usepackage{amsmath}
\usepackage{color}

\def\e{\begin{equation}}
\def\f{\end{equation}}

\title{Effects of Spatial Dispersion on Reflection from Mushroom-type Artificial Impedance Surfaces}

\author{Olli Luukkonen, M\'ario G.~Silveirinha, Alexander B.~Yakovlev, Constantin R.~Simovski, Igor S.~Nefedov, and Sergei A.~Tretyakov
\thanks{The work was supported in part by the Academy of
Finland through the Center-of-Excellence Program.}
\thanks{O.~Luukkonen, C.~R.~Simovski, I.~S.~Nefedov, and S.~A.~Tretyakov are with the Department of Radio Science and Engineering/SMARAD CoE - TKK Helsinki
University of Technology, P.O. 3000, FI-02015 TKK, Finland (email:
olli.luukkonen@tkk.fi),}
\thanks{M.~G.~Silveirinha is with Instituto
de Telecomunica\c c\~oes - Universidade de Coimbra, Departmento de
Engenharia Electrot\'ecnica, P\'olo II, 3030 Coimbra, Portugal.}
\thanks{A.~B.~Yakovlev is with the Department of Electrical Engineering - The University of Mississippi, University, MS 38677-1848, USA.}}

\begin{document}

\maketitle {\center \large }

\parskip 0pt

\begin{abstract}
Several recent works have emphasized the role of spatial dispersion
in wire media, and demonstrated that arrays of parallel metallic
wires may behave very differently from a uniaxial local material
with negative permittivity. Here, we investigate using local and
non-local homogenization methods the effect of spatial dispersion on
reflection from the mushroom structure introduced by Sievenpiper.
The objective of the paper is to clarify the role of spatial
dispersion in the mushroom structure and demonstrate that under some
conditions it is suppressed. The metamaterial substrate, or
metasurface, is modeled as a wire medium covered with an impedance
surface. Surprisingly, it is found that in such configuration the
effects of spatial dispersion may be nearly suppressed when the slab
is electrically thin, and that the wire medium can be modeled very
accurately using a local model. This result paves the way for the
design of artificial surfaces that exploit the plasmonic-type
response of the wire medium slab.


\end{abstract}

\section{Introduction}

Artificial impedance surfaces such as the corrugated surfaces (see
e.g. \cite{Kildal}) have been studied for decades. However, after
the seminal paper of D.~Sievenpiper et al. \cite{Sievenpiper} the
interest towards the artificial or {\itshape high-impedance}
surfaces boomed in the literature. The exotic features and the
possibility to engineer the response of the surface have resulted in
many novel and improved applications, such as quasi-TEM \cite{Yang}
and impedance waveguides \cite{Higgins,Luukkonen2}, band-gap
structures \cite{Rahmat-Samii}, low-profile antennas
\cite{feresidis,bell}, leaky-wave antennas
\cite{Sievenpiper_leaky1,Sievenpiper_leaky2}, and absorbers
\cite{gao,simms,engheta,Tretyakov_motl}, just to name a few.

In parallel with the research of new possibilities to utilize the
features of these exotic surfaces, also attempts to fully understand
the physics behind the surface behavior has been conducted. This
research has resulted in analytical formulas for different
artificial impedance surfaces providing physical insight and
engineering tools for the designers. In some cases the analyses are
based on the Floquet expansion of the scattered field
\cite{Goussetis,Maci}, whereas some models use extraction methods to
fit the model for surfaces. In other models the complete artificial
impedance surface structure is treated as a grounded slab of a
material with given permittivity and permeability tensors
\cite{Clavijo}. Although these models would predict the response of
the surfaces accurately, they are either laborious to use or lack
real physical basis.

In our previous work \cite{Luukkonen1} surfaces formed by an array
of rectangular patches over a grounded dielectric slab were
considered and an accurate spatially dispersive model for the patch
array was derived. This model for the patch array was later used in
the presence of a grounded dielectric slab perforated with metallic
vias \cite{Luukkonen2}. The perforated dielectric substrate was
modeled in the case of electrically thin substrates as a uniaxial
material with local negative permittivity. The corresponding
artificial dielectric designated here as wire medium is reasonably
well-known in the microwave engineering \cite{Brown,Rotman}, where
the wire medium was initially proposed to simulate electron plasma.
However, the wire medium has recently become known to exhibit
spatial dispersion \cite{Belov}. Non-local properties of the wire
medium have been used favorably, for instance, in subwavelength
imaging \cite{Belov_canal,Belov_lens,Ikonen_canal} or artificial
impedance surfaces \cite{mario1,mario2,mario3}.

In \cite{mario1,mario2,mario3} the grounded wire medium slab or
``Fakir's bed of nails'' has been studied taking the spatially
dispersive characteristics of the wire medium into account. These
works have demonstrated that the wire medium may behave very
differently from a uniaxial material with local negative
permittivity. It was proven that in the limit $a/h\rightarrow 0$
($h$ is the length of the wires and $a$ is the lattice constant) the
structured material behaves instead as a material with extreme
anisotropy with the relative permittivity along the wires
approaching to infinity and the relative transverse permittivity
equals unity. Based on these results, it is natural to ask if
spatial dispersion may also play an important role in the mushroom
structure, since it can be regarded as a wire medium capped with a
frequency-selective surface (FSS), as was done e.g. in
\cite{Clavijo,Luukkonen2,Luukkonen3}. In these works it was shown
that the response of the mushroom-type artificial impedance surface
may be predicted accurately by considering that the wire medium
behaves as a uniaxial material with local negative permittivity, in
apparent contradiction with the studies of
\cite{Belov,mario1,mario2,mario3}.

To study these issues, here we derive a homogenization model for the
mushroom structure that fully takes into account the effects of
spatial dispersion in the wire medium. Our objective in this paper
is to clarify the role of spatial dispersion in the mushroom
structure due to the aforementioned discrepancies in the treatment
of the wire medium. Surprisingly, our results show that unlike in
the topology for which the array of patches is removed (i.e. the
``Fakir's bed of nails'' studied in \cite{mario1}), the effect of
spatial dispersion may be nearly negligible in the mushroom
structure when the wire medium slab is electrically thin
\cite{Olli,Alexander,Yakovlev,Alexander2}. This result is in good
agreement with the results of \cite{Demetriadou}, where it was also
noticed that by connecting conductive structures, such as metal
plates, to the wires the spatial dispersion in wire medium can be
avoided. In such circumstances, the wire medium may be modeled to a
good approximation as a uniaxial material with local negative
permittivity. We demonstrate that the effects of spatial dispersion
are significant in mushroom structures only for rather thick wire
medium slabs.

The rest of the paper is organized as follows: First, we will
discuss and derive two analytical models of the electromagnetic
response of the mushroom structure. One of the models takes into
account the effect of spatial dispersion in the wire medium, while
the other model assumes that the wire medium has a local response.
Based on the homogenization results, we will discuss in which
circumstances the spatial dispersion effects are suppressed. Then,
in section III, we will validate the two models against full wave
simulations, demonstrating that by taking into account spatial
dispersion, we obtain an accurate model for the response of the
mushroom structures. However, under some circumstances, that are
well applicable to most of the practical high-impedance surfaces,
the simple model neglecting the spatial dispersion in the wire
medium is accurate and perfectly capable of predicting the response
of the surface to incident plane waves.


\section{Analytical models for the mushroom structure}

\begin{figure}[t!]
\centering
\subfigure[]{\includegraphics[width=4.5cm]{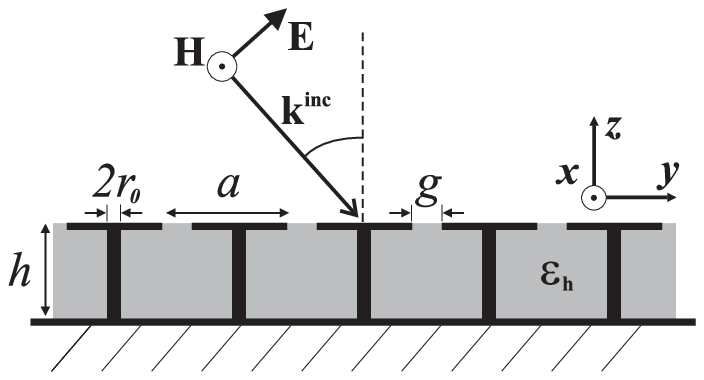} }
\subfigure[]{\includegraphics[width=4cm]{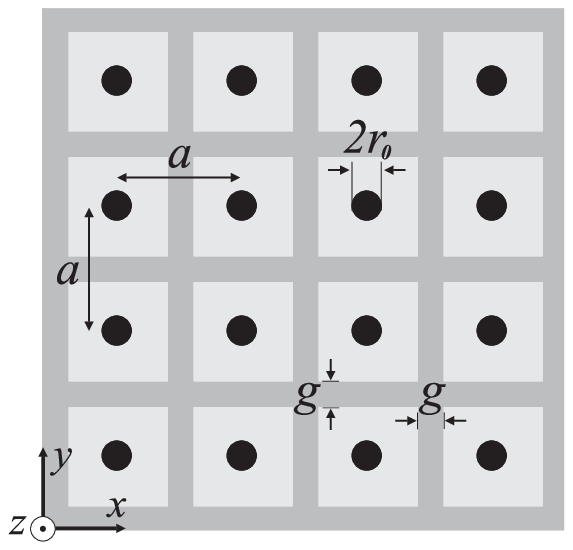}}
\caption{Illustration of the mushroom structure (a) from the side
and (b) from the above. The structure comprises a patch array over a
dielectric slab perforated with metallic vias. The periodicity of
the vias and the patches is $a$, the gap between the adjacent
patches $g$, the radius of the vias $r_0$, and the relative
permittivity of the host medium $\varepsilon_{\rm h}$.} \label{fig1}
\end{figure}

A schematic picture of the mushroom structure comprising an array of
patches over a dielectric layer perforated with metallic vias is
shown in Fig.~\ref{fig1}. Actually, the models proposed in this work
are also applicable for other topologies of the mushroom structure,
as will be discussed later. Here the patch array serves merely as a
good example for its spatially dispersive electromagnetic properties
have been studied in our previous work \cite{Luukkonen1}. The
surface impedance describing the electromagnetic properties of the
square patch array reads \cite{Luukkonen1}: \e Z_{\rm g}^{\rm TM} =
-j\frac{\eta_{\rm eff}}{2\alpha}, \label{eq:Z_g}\f with $\eta_{\rm
eff} = \sqrt{\mu_0/\varepsilon_0\varepsilon_{\rm eff}}$ and the grid
parameter $\alpha$ is given by: \e \alpha = \frac{k_{\rm
eff}a}{\pi}\ln\left(\sin^{-1}\left(\frac{\pi g}{2a} \right)
\right),\f where $k_{\rm eff}=k_0\sqrt{\varepsilon_{\rm eff}}$, $a$
is the period of the array, $g$ is the gap between the adjacent
patches, and $k_0$ is the wave number in free space. Furthermore,
the effective relative permittivity reads: \e \varepsilon_{\rm eff}
= \frac{\varepsilon_{\rm h} + 1}{2},\f where $\varepsilon_{\rm h}$
is the relative permittivity of the supporting host medium.

We model the dielectric slab perforated with metallic vias as wire
medium which can be described with an effective relative
permittivity tensor having values only along its diagonal. Following
the notations of Fig.~\ref{fig1}, the permittivity tensor for this
particular case reads $\overline{\overline{\varepsilon}}_{\rm r,
eff}= \varepsilon_{\rm h}\left( \mathbf{\hat{u}}_{\rm
x}\mathbf{\hat{u}}_{\rm x} + \mathbf{\hat{u}}_{\rm
y}\mathbf{\hat{u}}_{\rm y}\right) + \varepsilon_{\rm
zz}\mathbf{\hat{u}}_{\rm z}\mathbf{\hat{u}}_{\rm z}$, where
$\varepsilon_{\rm h}$ is the relative permittivity of host medium
and $\varepsilon_{\rm zz}$ is the effective relative permittivity
along the vias. In what follows, we describe two analytical models
for the electromagnetic response of the considered structured
substrate. The main difference between the two models is related to
how the wire medium slab is treated. The second model, described in
sub-section B, considers only the effects of frequency dispersion in
the wire medium, and assumes that the material has a local response.
However, it is known that the wire medium suffers from spatial
dispersion even at very low frequencies \cite{Belov}. Also it is
known that for an electrically thin wire medium slab (with no patch
array) the charges accumulate on the tips of the metallic vias and
the effects of spatial dispersion need to be taken into account
\cite{mario1}. For this reason, in sub-section A, we describe a
non-local model that takes both TM- (with respect to $z$) and
TEM-polarized waves into account in the wire medium slab. In this
case we assume that the effective relative permittivity of the wire
medium slab along the vias has both frequency and spatial
dispersion.

\subsection{Non-local model for the wire medium}

For long wavelengths the relative effective permittivity of the wire
medium along the metallic vias (normal direction) can be written for
the non-local model as \cite{Belov}: \e \varepsilon_{\rm zz} =
\varepsilon_{\rm h}\left(1 - \frac{k_{\rm p}^2}{k^2 - q_{\rm
z}^2}\right), \label{eq:epsilon_zz_nonlocal}\f with
$k=k_0\sqrt{\varepsilon_{\rm h}}$ is the wave number in the host
medium, $q_{\rm z}$ is the z-component of the wave vector
$\mathbf{q} = (q_{\rm x},q_{\rm y},q_{\rm z})$, and $k_{\rm p}$ is
the plasma wave number given as \cite{Belov}: \e \left(k_{\rm
p}a\right)^2 = \frac{2\pi}{\ln\left( \frac{a}{2\pi r_0}\right) +
0.5275}, \f where $a$ is the period of the vias (the same period
with the patch array) and $r_0$ is the radius of the vias (see
Fig.~\ref{fig1}).

A TM-polarized incident plane wave excites  TEM- and TM-polarized
plane waves in the wire medium slab through the capacitive array.
For these two plane waves we have the following dispersion
equations: \e k = \pm q_{\rm z}, \hspace{1cm} \textrm{(TEM mode)}
\label{eq:k_TEM_polarization}\f \e k^2 = k_{\rm p}^2 +
\mathbf{q}\cdot\mathbf{q}. \hspace{1cm}
\label{eq:k_TM_polarization}\textrm{(TM mode)} \f

In \cite{mario1,mario2} the grounded wire medium slabs were
characterized by solving the field amplitudes in all space. Omitting
the dependency on the $y$-coordinate ($e^{-jk_{\rm y}y}$), we can
write the field amplitudes for the magnetic fields similarly as: \e
H_{\rm x}= \left\{\begin{array}{ll}
e^{jk_{\rm z}z} + \rho e^{-jk_{\rm z}z}, & \textrm{if $z>0$}\\
A_{\rm TEM}\cos\left(k(z+h) \right) + \\
\hspace{0.3cm}+A_{\rm TM} \cosh\left(\gamma_{\rm TM}(z+h) \right), &
\textrm{if $-h<z<0$}\end{array} \right. \label{eq:magneticfields}\f
where $k_{\rm z}=\sqrt{k_0^2 - k_{\rm t}^2}$, $\gamma_{\rm TM} =
\sqrt{k_{\rm p}^2 + k_{\rm t}^2 - k^2}$, and $k_{\rm t}$ is the
transverse wave number, which is determined by the angle of
incidence. In order to solve for the unknowns $\rho$, $A_{\rm TEM}$,
and $A_{\rm TM}$, an additional boundary condition (ABC) is needed.
In \cite{mario3} an ABC was derived for the interface between air
and the wire medium. This ABC is not applicable in the present
problem, because the patch array is in galvanic connection with the
wires. However, in \cite{mario2} an ABC that properly models the
connection between the wire medium and a metallic surface was
derived, and such ABC is applicable also in our case. It should be
noticed here, however, that this boundary condition assumes that the
charge density vanishes at the connection point of the metallic wire
and the element of the capacitive array. This condition does not
always hold; for instance in cases where the array element size is
of the same order as the diameter of the wire. Anyway, for
high-impedance surfaces, such as the mushroom structure studied
here, this is seldom the case, because miniaturization of the
surfaces leads inevitably to large array element size with respect
to the period of the array (and to the period of the wire medium).
In our case the additional boundary condition reads on the wire
medium side of the patch array ($z=0^-$) as \cite{mario2}:
\begin{eqnarray} \frac{d}{dz}\left(\omega\varepsilon_0\varepsilon_{\rm
h}\mathbf{\hat{u}_{\rm z}}\cdot\mathbf{E}
+\left(\mathbf{\hat{u}_{\rm z}}\times \mathbf{k_{\rm
t}}\right)\cdot\mathbf{H} \right)=0 \nonumber\\
\Rightarrow k_0\varepsilon_{\rm h} \frac{dE_{\rm z}}{dz}- k_{\rm
t}\eta_0\frac{dH_{\rm x}}{dz} = 0. \label{eq:ABC}
\end{eqnarray}

As in \cite{mario3}, the remaining boundary conditions are obtained
from the classical boundary conditions with the exception that in
our case the transverse magnetic field is discontinuous over the
patch array. So, at the interface between the wire medium and air
($z=0$) we have the following conditions of continuous tangential
electric and discontinuous magnetic field, respectively: \e
\left.E_{\rm y}\right|_{\rm z=0^+} - \left.E_{\rm y}\right|_{\rm
z=0^-} = 0, \label{eq:continuity}\f \e \left.H_{\rm x}\right|_{\rm
z=0^+} - \left.H_{\rm x}\right|_{\rm z=0^-} = Z_{\rm g}^{-1}E_{\rm
y}. \label{eq:discontinuity}\f  Here, $Z_{\rm g}$ is the surface
impedance of the patch array given by \eqref{eq:Z_g}. In
\eqref{eq:ABC}--\eqref{eq:discontinuity} the values for the $z$- and
$y$-components of the electric field are calculated from
\eqref{eq:magneticfields} by using Maxwell's equations, taking into
account that the material is nonlocal and that the permittivity
tensor depends on the considered mode.

Finally, using \eqref{eq:ABC}--\eqref{eq:discontinuity} we can solve
the reflection coefficient $\rho$ (for the magnetic field) from
\eqref{eq:magneticfields} unambigiously: \e \rho =
\frac{\frac{\varepsilon_{\rm zz}^{\rm TM}}{\gamma_{\rm
TM}}\coth\left(\gamma_{\rm TM}h\right) + \frac{\varepsilon_{\rm
zz}^{\rm TM} - \varepsilon_{\rm h}}{k}\cot\left(kh\right) +
\frac{\eta_0}{jk_0}Z_{\rm g}^{-1} - \frac{1}{jk_{\rm
z}}}{\frac{\varepsilon_{\rm zz}^{\rm TM}}{\gamma_{\rm
TM}}\coth\left(\gamma_{\rm TM}h\right) + \frac{\varepsilon_{\rm
zz}^{\rm TM} - \varepsilon_{\rm h}}{k}\cot\left(kh\right) +
\frac{\eta_0}{jk_0}Z_{\rm g}^{-1} + \frac{1}{jk_{\rm z}}},
\label{eq:rho} \f where the relative effective permittivity along
the direction of the metallic vias is written for the TM
polarization using \eqref{eq:epsilon_zz_nonlocal} and
\eqref{eq:k_TM_polarization} as: \e \varepsilon_{\rm zz}^{\rm TM} =
\varepsilon_{\rm h}\left(1-\frac{k_{\rm p}^2}{k_{\rm p}^2 + k_{\rm
t}^2}\right).\f The reflection coefficient $\rho$ given in
\eqref{eq:rho} can be used to characterize the electromagnetic
properties of the high-impedance surface in the case of TM-polarized
plane wave excitation. For a TE-polarized incident wave the
transverse electric field does not excite the metallic vias. In this
case the models presented in \cite{Luukkonen1} can be applied for
modeling of the response of the surface. Furthermore, the present
model is not only applicable for this particular type of
high-impedance surface structure but for other type of surfaces as
well, as long as the structure consists of a wire medium slab. In
these cases $Z_{\rm g}$, responsible for the interaction with the
capacitive array, should be recalculated appropriately.

Using \eqref{eq:ABC}--\eqref{eq:rho} we can solve also
\eqref{eq:magneticfields} for the amplitudes of the magnetic fields
inside the high-impedance surface structure. We use them to
determine the microscopic current on the metallic wires and use this
quantity in order to validate our assumptions on the current phase
variation in the case of the local model for the wire medium, as
will be discussed ahead. The averaged current along the metallic
wires can be written in terms of averaged macroscopic fields and the
microscopic current, $I$, as (see \cite{mario2}): \e J_{\rm av, z} =
\frac{1}{a^2}I = -j\omega\varepsilon_0\varepsilon_{\rm h}E_{\rm
z}+jk_{\rm t}H_{\rm x}. \f

\subsection{Local model for the wire medium}

In this model for the high-impedance surface we assume that the
effective relative permittivity of the wire medium slab along the
direction of the metallic vias can be described using the local
model: \e \varepsilon_{\rm zz}^{\rm loc} = \varepsilon_{\rm
h}\left(1 - \frac{k_{\rm p}^2}{k_0^2\varepsilon_{\rm h}}\right)
\label{eq:epsilon_zz_local} \f In this approximation we have assumed
the current phase variation along the wires to be minimum and the
patch array over the wire medium slab to operate as a nearly-perfect
reflector. Because of the two mirror boundaries the wires appear
infinite for the electromagnetic wave inside the grounded wire
medium slab. In addition, the minimum current phase variation along
the vias suggests that the waves can propagate mainly along the
transverse direction inside the wire medium ($q_{\rm z} \approx 0$),
which makes, together with the infinite vias, our quasi-static
approximation justified. Within this model the incident plane wave
excites a single mode (the extraordinary wave) inside the wire
medium. It should be emphasized that in general the properties of
this extraordinary mode have nothing to do with the properties of
either the TEM or TM waves predicted by the non-local model. This
model that treats the wire medium slab using the local approximation
for the effective permittivity will be referred to as
\textit{epsilon-negative} (ENG) model from here on in order to
distinguish it from the model derived in the previous section. As
discussed in the previous section, that model uses non-local
(spatially dispersive) approximation for the effective permittivity
and requires the use of additional boundary conditions. For this
reason the model presented in the previous section is referred to as
\textit{spatially dispersive} (SD) model from here on.

A simple but yet accurate analytical model for the mushroom
structure using this local approximation of permittivity of the wire
medium slab has been derived in our previous work \cite{Luukkonen2}.
In order to compare the results of this paper with our previous
results we rewrite the results of \cite{Luukkonen2} in a form
similar to \eqref{eq:rho}. After some algebra the results of
\cite{Luukkonen2} can be rewritten for the magnetic field reflection
coefficient in the following form: \e \rho_{\rm loc} =
\frac{\frac{\varepsilon_{\rm h}}{\gamma_{\rm
loc}}\coth\left(\gamma_{\rm loc}h \right)+\frac{\eta_0}{jk_0}Z_{\rm
g}^{-1} - \frac{1}{jk_{\rm z}}}{\frac{\varepsilon_{\rm
h}}{\gamma_{\rm loc}}\coth\left(\gamma_{\rm loc}h
\right)+\frac{\eta_0}{jk_0}Z_{\rm g}^{-1} + \frac{1}{jk_{\rm z}}},
\label{eq:rho_loc}\f where the propagation constant along the
$z$-axis in the local approximation is given as \e \gamma_{\rm loc}
= \sqrt{\frac{k_{\rm t}^2}{\varepsilon_{\rm zz}^{\rm loc}} -
k_0^2\varepsilon_{\rm h}} \label{eq:gamma_loc}\f

When comparing the SD model with the ENG model, that is
Eqs.~\eqref{eq:rho} and \eqref{eq:rho_loc}, we find that the
difference between the two models lies on how they treat the surface
impedance of the grounded wire medium slab, as stated earlier in
this paper. For the SD model and the ENG model the normalized
surface admittances read, respectively: \e y_{\rm s, SD} =
\frac{\varepsilon_{\rm zz}^{\rm TM}}{\gamma_{\rm
TM}}\coth\left(\gamma_{\rm TM}h\right) + \frac{\varepsilon_{\rm
zz}^{\rm TM} - \varepsilon_{\rm h}}{k}\cot\left(kh\right),
\label{eq:y_s_SD}\f \e y_{\rm s, ENG} = \frac{\varepsilon_{\rm
h}}{\gamma_{\rm loc}}\coth\left(\gamma_{\rm loc}h \right)
.\label{eq:y_s_ENG}\f The surface admittance is given in terms of
the normalized surface admittance as
$Y_{\rm s} = jy_{\rm s}k_0/\eta_0$. 

\section{Numerical validation}

In order to compare the model derived in this paper, \eqref{eq:rho},
and the result of our previous work, \eqref{eq:rho_loc}, we use
these models to characterize two particular high-impedance surfaces.
We will do the characterization in terms of reflection phase
diagrams for the electric field. We will also validate the results
by numerical simulations obtained with Ansoft's High Frequency
Structure Simulator (HFSS) \cite{HFSS} and with CST Microwave Studio
\cite{CST}. Following the notations in Fig.~\ref{fig1}, the
parameters for the two cases read ($r_0=0.05$\,mm in both cases):
\vspace{0.1cm}
\begin{itemize}
\item[1)] $a = 2$\,mm, $g=0.2$\,mm, $h=1$\,mm, and
$\varepsilon_{\rm h} = 10.2$,\vspace{0.1cm}
\item[2)] $a = 1$\,mm, $g=0.1$\,mm, $h=5$\,mm, and
$\varepsilon_{\rm h} = 1$
\end{itemize} \vspace{0.1cm}

\begin{figure}[t!]
\centering
\subfigure[]{\includegraphics[width=0.5\textwidth]{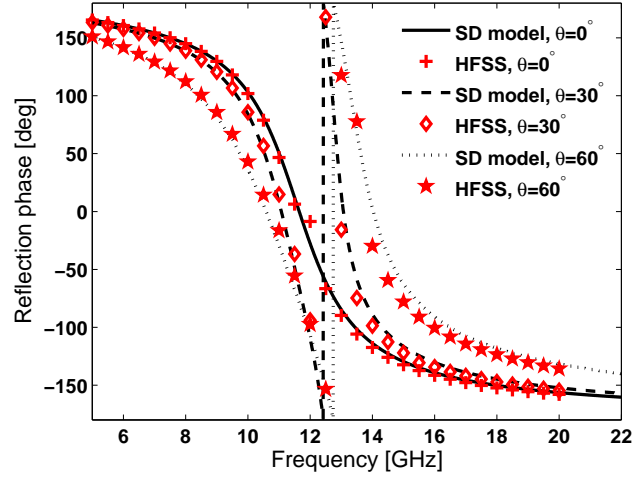} }
\subfigure[]{\includegraphics[width=0.5\textwidth]{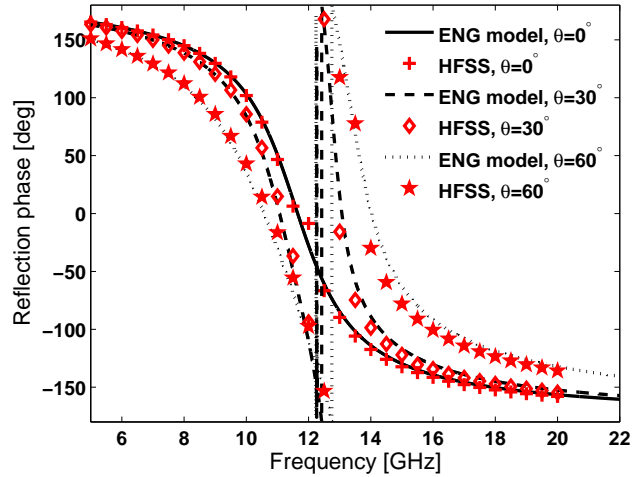}}
\caption{The reflection phase calculated using (a) SD model in
\eqref{eq:rho} and (b) ENG model in \eqref{eq:rho_loc} for the first
example: $a=2$\,mm, $g=0.2$\,mm, $h=1$\,mm, $\varepsilon_{\rm r}
=10.2$, and $r_0=0.05$ \,mm.} \label{fig:rphase_low}
\end{figure}

In Figs.~\ref{fig:rphase_low}(a) and (b) the reflection phase
according to the SD and ENG models, respectively, for the first
example are given. Quite interestingly, the general agreement
between the two models for this example is excellent,
notwithstanding the fact that each model describes the wire medium
slab with very different material parameters. It is also seen that
both models agree very well with the HFSS results. In the absence of
losses, the ENG model shows some spurious resonances on a very
narrow frequency band in the close vicinity of the plasma frequency
of the wire medium ($f_{\rm p}/\sqrt{\varepsilon_{\rm
h}}=12.1$\,GHz). The spurious resonances appear due to the fact that
in the close vicinity of the plasma frequency $\varepsilon_{\rm
zz}\rightarrow0$ which according to \eqref{eq:gamma_loc} invalidates
our assumption on an electrically thin substrate. The issue of
spurious resonances has been discussed in more detail in
\cite{Luukkonen3}. The results of Fig.~\ref{fig:rphase_low} suggest
that for the considered geometry the effects of spatial dispersion
are negligible. It should be emphasized that this characteristic is
radically different from the results obtained for a grounded wire
medium slab with no patch array, for which, as demonstrated in
\cite{mario1}, the wire medium slab has a completely different
electromagnetic response.

Outside of the very narrow frequency band of the spurious
resonances, both models show also an additional physical resonance
for oblique incidence. This resonance occurs due to epsilon-near
zero value of the $\varepsilon_{\rm zz}$.

To further understand the physical reasons why the effects of
spatial dispersion are negligible for this example, we have plotted
in Fig.~\ref{fig:currentmagn}\,(a) the magnitude of the normalized
current profile along the vias. It can be seen that the current
varies very little along the wires. The phase of the current (not
reported here for brevity) is also practically constant.  This
behavior of the electric current is consistent with the hypotheses
used to derive the ENG model, and in particular imply that the
electromagnetic fields below the patch grid are nearly uniform along
$z$, i.e. $\partial/\partial z \approx 0$. Thus, in the spectral
(Fourier) domain the electromagnetic field amplitude in the wire
medium has a peak at $q_{\rm z} \approx 0$. This explains the
suppression of spatial dispersion, because the nonlocal dielectric
function \eqref{eq:epsilon_zz_nonlocal} is coincident with the local
dielectric function \eqref{eq:epsilon_zz_local} when $q_{\rm z}=0$.

\begin{figure}[t!]
\centering \includegraphics[width=0.5\textwidth]{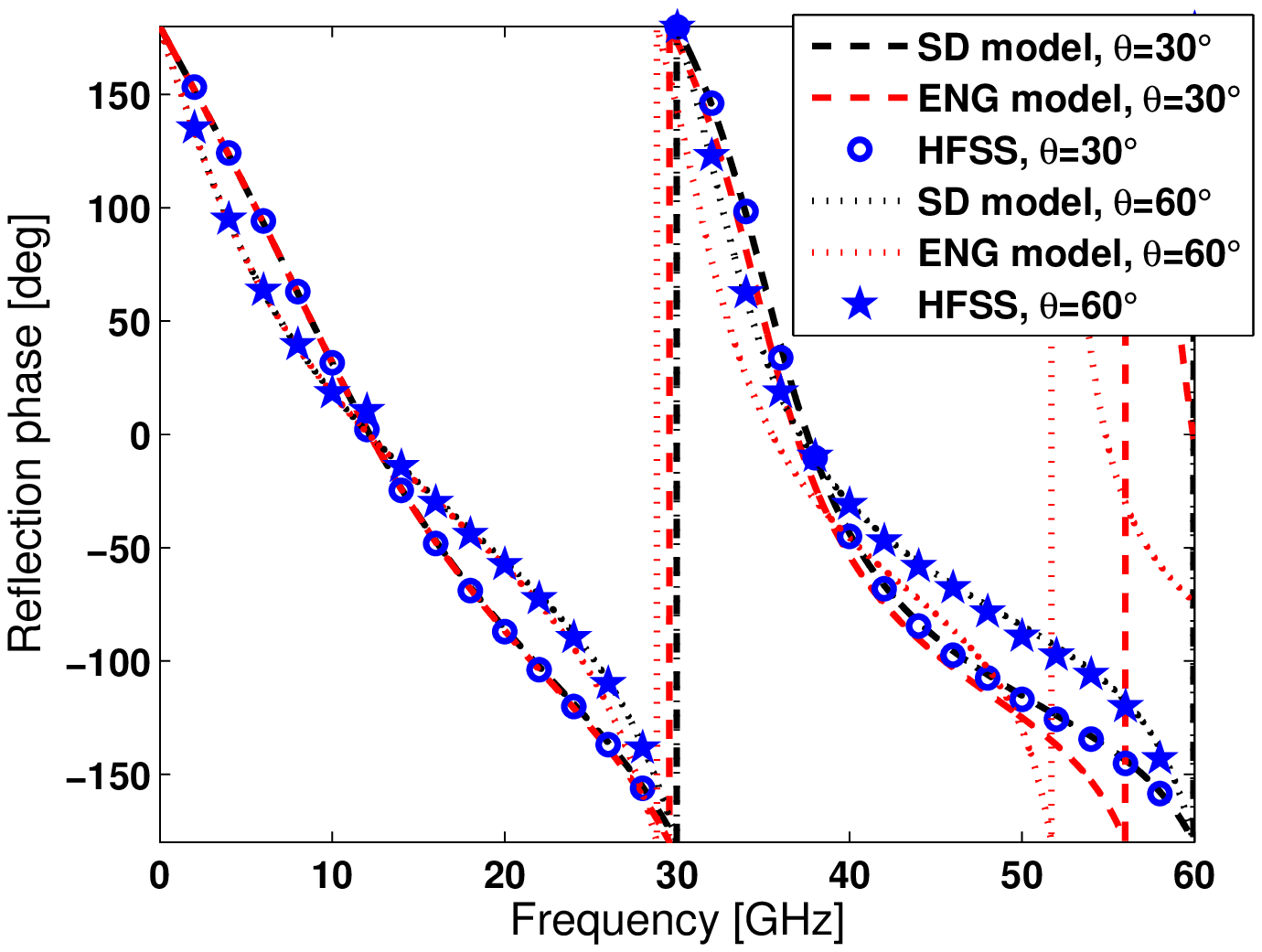}
\caption{The reflection phase calculated using the SD model in
\eqref{eq:rho} and the ENG model in \eqref{eq:rho_loc} for the
second example: $a=1$\,mm, $g=0.1$\,mm, $h=5$\,mm, $\varepsilon_{\rm
r} =1$, and $r_0=0.05$ \,mm.} \label{fig:rphase_high}
\end{figure}

\begin{figure}[t!]
\centering \mbox{
\subfigure[]{\includegraphics[width=0.24\textwidth]{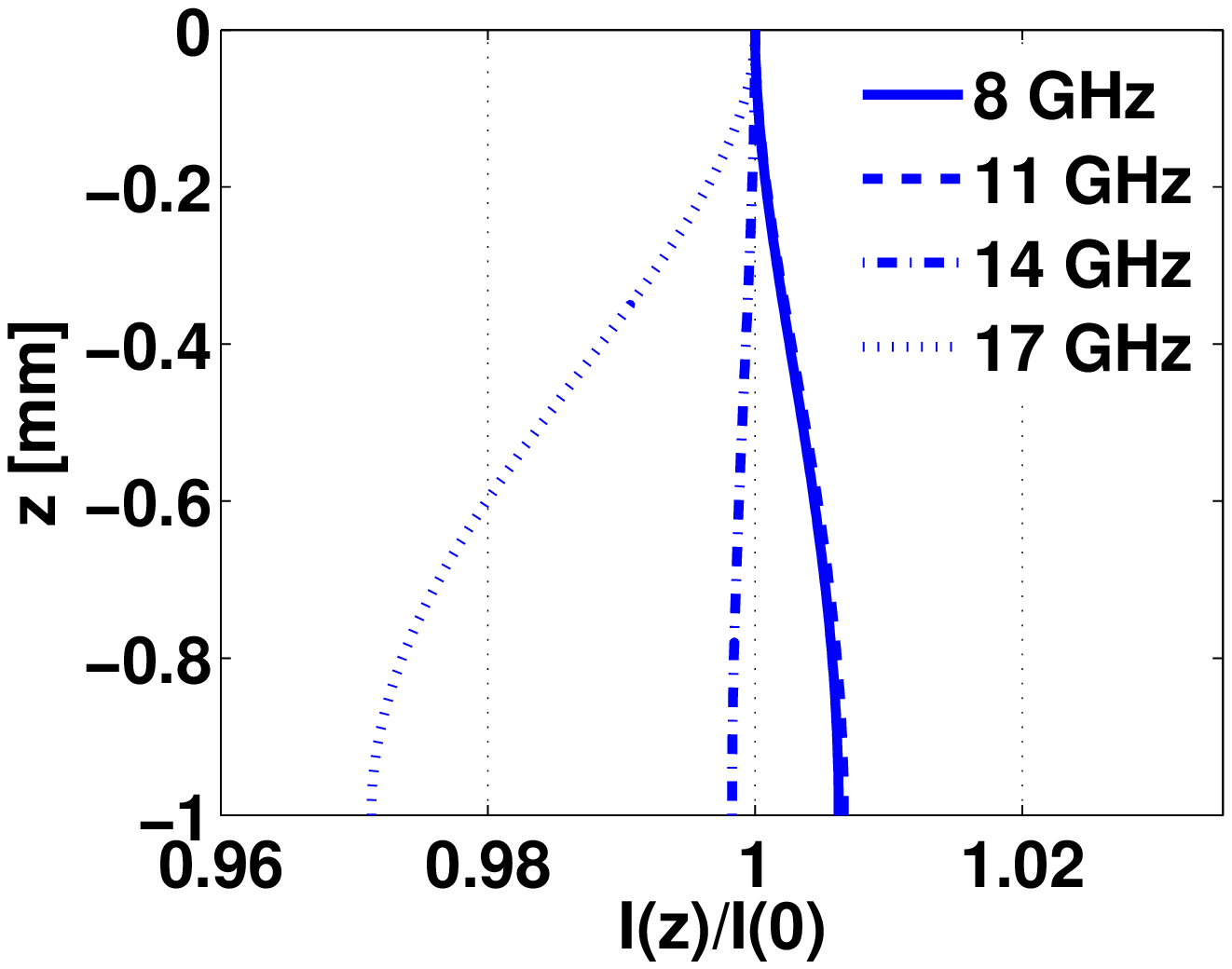}
}
\subfigure[]{\includegraphics[width=0.24\textwidth]{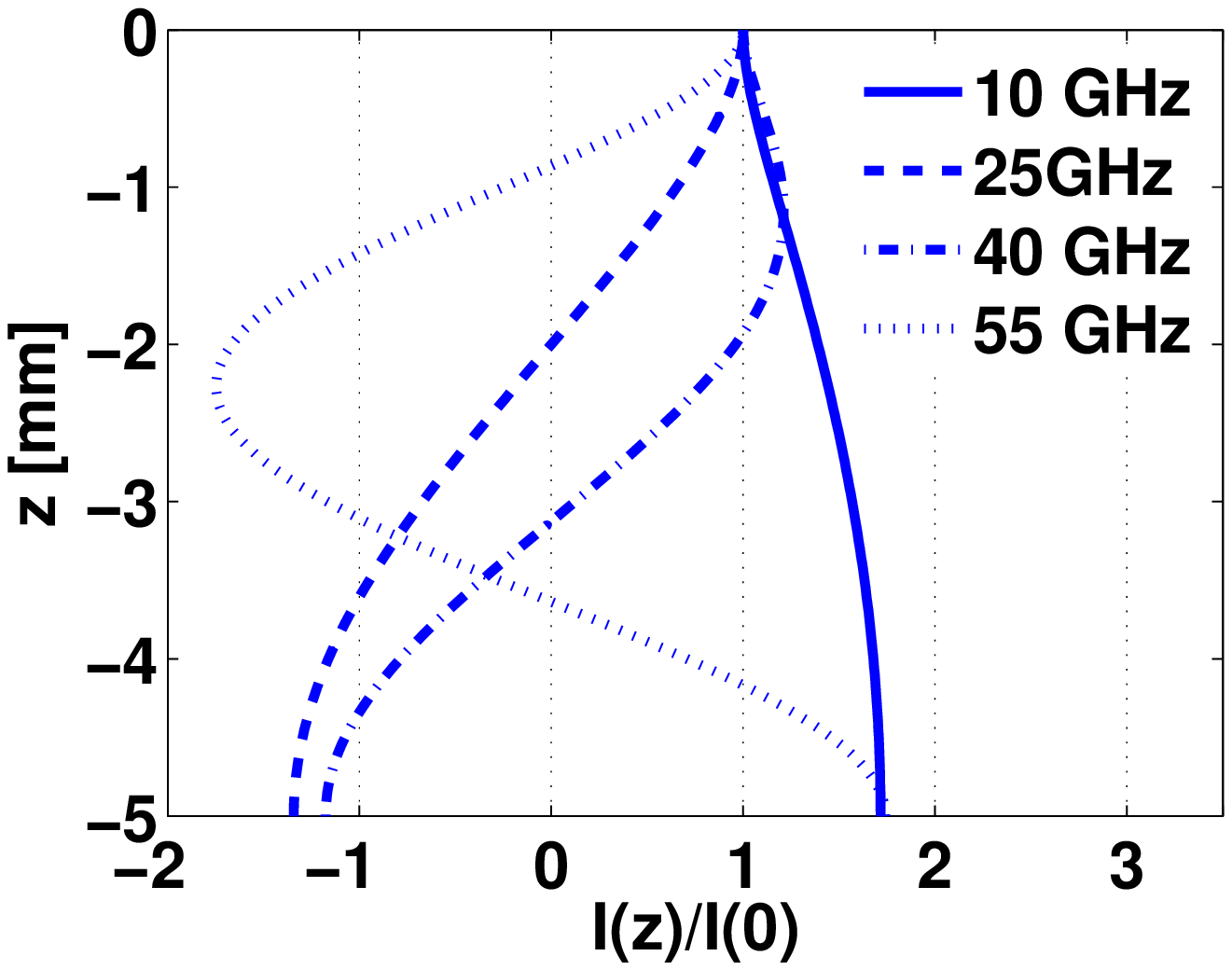}}}
\caption{The current magnitudes at different frequencies (a) for the
first example structure and (b) for the second example structure
normalized to the current value on the patch array ($h=0$). The
reader should notice the different frequencies and scales in the
plots. The angle of incidence is 30 degrees for all cases.}
\label{fig:currentmagn}
\end{figure}

In Fig.~\ref{fig:rphase_high} the reflection phase obtained with the
SD and ENG models for the second example are compared against the
simulation results. The models are in good agreement with each other
below 20\,GHz. The disagreement between the models becomes more
noticeable as the frequency increases. The HFSS simulation results
are in very good agreement with the results of the SD model through
the whole frequency band. If we look at the normalized current
magnitudes plotted in Fig.~\ref{fig:currentmagn}\,(b) for the
incidence angle of 30 degrees we see that at 10\,GHz the current
magnitude changes somewhat. In this case the current phase remains
constant. However, at 25\,GHz (the slab is electrically thicker) the
current magnitude changes drastically and the current phase is no
longer constant. This causes the charges to accumulate on the wires
and the effects of spatial dispersion are no longer negligible.

\begin{figure}[t!]
\centering
\subfigure[]{\includegraphics[width=0.5\textwidth]{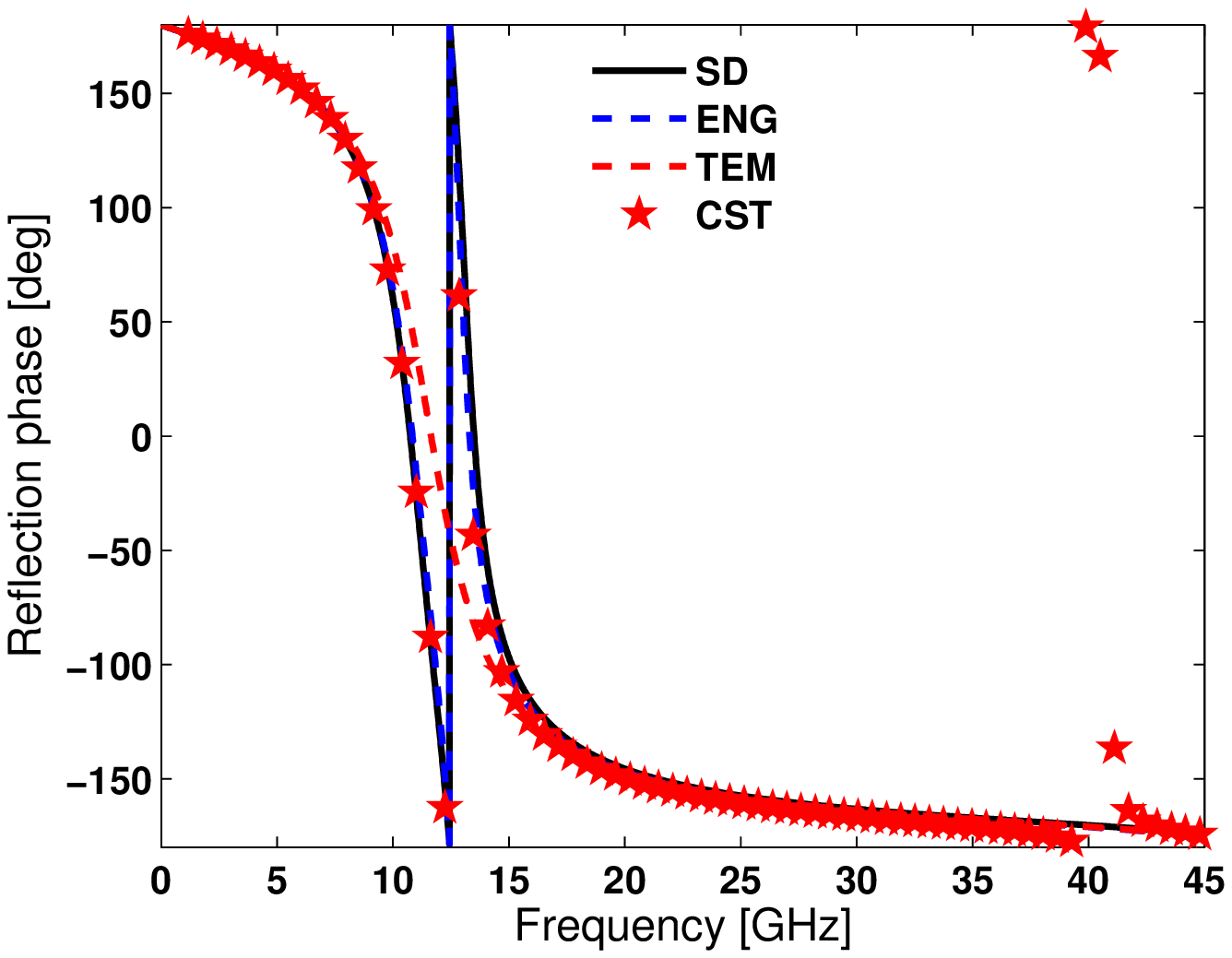} }
\subfigure[]{\includegraphics[width=0.5\textwidth]{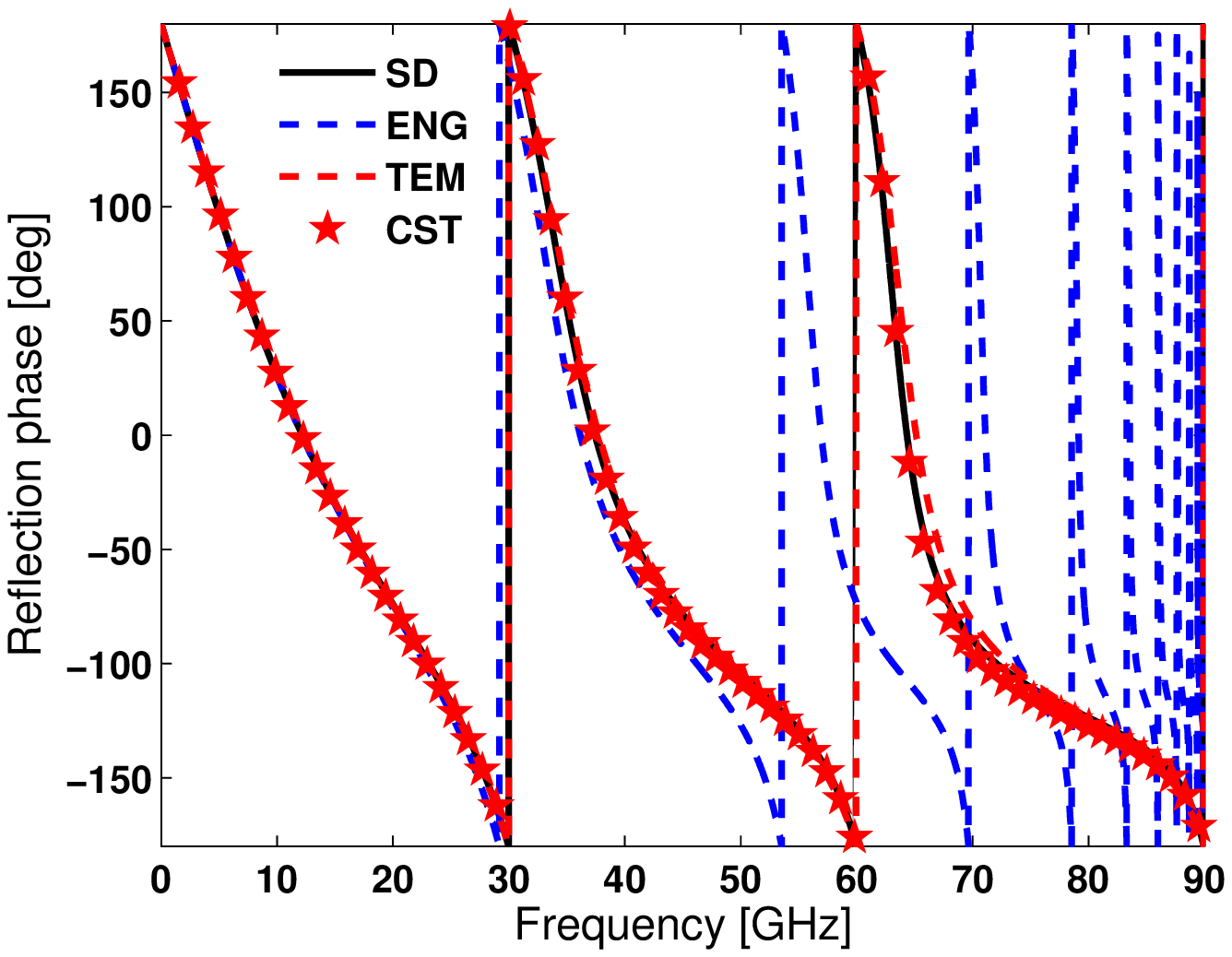}}
\caption{The reflection phase for (a) the second and (b) the first
example for the incidence angle of $45^\circ$. The reflection phases
according to the SD, ENG, and TEM (see \cite{Tretyakov})
approximations (plotted with solid black, dashed blue, and red
curves, respectively) are compared with each other and against
CST-simulation results (plotted with red stars). The parameters for
the structures are given on page 4.} \label{fig:TEMcomparison}
\end{figure}

The results in Figs.~\ref{fig:rphase_low} and \ref{fig:rphase_high}
suggest that the SD and ENG models give the same results when the
current along the vias is essentially constant, or equivalently when
the wire medium slab is electrically thin ($kh\ll1$) and the height
of the metallic pins is small as compared to the lattice constant
$h/a<1$. Indeed, it may be verified by a straightforward analysis,
that under these assumptions the normalized surface admittances in
\eqref{eq:y_s_SD} and \eqref{eq:y_s_ENG} reduce to: \e y_{\rm s, SD}
\approx \frac{\varepsilon_{\rm zz}^{\rm TM}}{\gamma_{\rm TM}^2h} +
\frac{\varepsilon_{\rm zz}^{\rm TM} - \varepsilon_{\rm
h}}{k^2h}=\frac{\varepsilon_{\rm h}}{\gamma_{\rm loc}^2h},\f \e
y_{\rm s, ENG} \approx \frac{\varepsilon_{\rm h}}{\gamma_{\rm
loc}^2h},\f respectively. Here we have used the fact that $k_{\rm
p}\sim 1/a$.

Besides the case in which the current along the wires is nearly
uniform, the effects of spatial dispersion may also be negligible in
the regime where the wire medium is characterized by extreme
anisotropy. This situation is illustrated in
Fig.~\ref{fig:rphase_high}, where, notwithstanding the current
varying drastically along the wires, the agreement between the ENG
and the SD models is good.  These results are explained by the fact
that we operate well below the plasma frequency of the wire medium.
Indeed, if we assume $k_{\rm p}\gg k$ we have for the relative
effective permittivity along the wires in the non-local model for
the TM polarization $\varepsilon_{\rm zz}^{\rm TM} \approx 0$ and in
the local model for the extraordinary mode $\left|\varepsilon_{\rm
zz}^{\rm loc} \right| \rightarrow \infty\Rightarrow \gamma_{\rm loc}
= jk_0\sqrt{\varepsilon_{\rm h}}$. Under these assumptions the
normalized admittances in \eqref{eq:y_s_SD} and \eqref{eq:y_s_ENG}
both reduce to: \e y_{\rm s, SD} \approx y_{\rm s, ENG} \approx -
\frac{\varepsilon_{\rm h}}{k}\cot\left(kh\right),
\label{eq:admittance_TEM}\f which, in fact, corresponds to the
normalized admittance according to the TEM model \cite{Tretyakov}.
The above result proves that the reflection properties predicted by
the two analytical models are indeed the same under the regime of
extreme anisotropy.

However, when these conditions (i.e. uniform current or extreme
anisotropy) are not observed, the response of the mushroom
structures may be dominated by non-local effects. To illustrate
this, we plot in Fig.~\ref{fig:TEMcomparison} the reflection phase
calculated with the SD and ENG models for the same two examples
considered before, and an angle of incidence of $45^\circ$. The
star-shaped symbols in Fig.~\ref{fig:TEMcomparison} correspond to
full-wave results obtained with the commercial simulator CST
Microwave Studio \cite{CST}. As can be seen, in this example where
$h/a\gg1$ the SD results concur very well with full-wave
simulations, whereas the ENG model only gives meaningful results for
frequencies well below the plasma frequency, where, as discussed
before the material is characterized by extreme anisotropy and
spatial dispersion is suppressed. Quite interestingly, in the regime
where $h/a\gg1$ the response of the structure is dominated by the
effects of the TEM mode, consistent with the results of
\cite{Ikonen_canal}. The TEM approximation corresponds to the case
where the only mode excited in the wire medium is the TEM mode, i.e.
the TM mode is completely discarded when calculating the response
from the wire medium slab. According to the analysis in
\cite{Tretyakov}, when $h/a\gg1$, the TM mode in the wire medium is
nearly suppressed. To confirm this, we have plotted in
Fig.~\ref{fig:TEMcomparison} the results obtained under such TEM
approximation. These results are obtained by considering that
$A_{TM}=0$ in \eqref{eq:magneticfields} and imposing only the
boundary conditions \eqref{eq:continuity}--\eqref{eq:discontinuity}
at the interface $z=0$. The results of
Fig.~\ref{fig:TEMcomparison}~(b) (the plasma frequency, $f_{\rm
p}/\sqrt{\varepsilon_{\rm h}}\approx92.1$\,GHz) show that, in fact,
the TEM approximation agrees well with the SD model and with the
simulation results whereas the ENG model begins to disagree with the
simulated results when the electrical thickness of the slab becomes
comparable with the wavelength. However, the results in
Fig.~\ref{fig:TEMcomparison}(a) for the first example of mushroom
structure ($f_{\rm p}/\sqrt{\varepsilon_{\rm h}} \approx 12.1$\,GHz)
show that in the vicinity of the plasma frequency the TEM model
fails, as expected, whereas the SD and ENG models fit well with the
CST simulations (until the limit $ka=\pi$, after which the
homogenization fails). Indeed, the TEM approximation is suitable
only for frequencies well below the plasma frequency.

\section{Discussion and conclusions}

In this paper the reflection characteristics of mushroom-type
high-impedance surface were studied. Two analytical models, namely
ENG and SD models, were investigated for the electromagnetic
response of high-impedance surfaces formed by a capacitive array
over a grounded dielectric slab perforated with metallic pins. We
derived the novel analytical (SD) model for the mushroom structure
that takes the spatially dispersive characteristics of the grounded
wire medium slab into account. Using this and a more simple model
from our previous work (ENG model), we studied the effect of the
spatial dispersion in the wire medium on the reflection properties
of the mushroom structure.

Surprisingly, our results demonstrate that the two models concur
very well in a regime where the thickness of the slab is much
smaller than the radiation wavelength, or alternatively when the
wire medium is characterized by extreme anisotropy. In these
conditions, the wire medium slab may, indeed, be regarded as a
uniaxial material with negative permittivity. This property
contrasts markedly with other wire medium topologies, for which the
effects of spatial dispersion are dominant \cite{Belov_lens,mario1}.
This dramatically different behavior is explained by the fact that
when the wires are connected to the metallic patches/ground plane,
the charges no longer accumulate at the tips of the metallic vias
and do not create the additional current flow on the vias which
would cause the spatially dispersive effects. Hence, the current
along the wires can be nearly uniform and as a consequence the
polarization along the wires is out of phase with the macroscopic
electric field, as in an electron plasma.

Furthermore, the assumptions behind the ENG model are verified by
the results of this paper. We validated the novel model using
full-wave simulations, showing that it is very accurate even when
the current amplitude and the phase vary considerably. We also
discussed the validity limits of the ENG model. It can be concluded
that the spatial dispersion in the wire medium of the mushroom-type
high-impedance surface structure can be suppressed in particular
designs, i.e. when the patch size is sufficiently large and the
mushroom structure is electrically thin. This applies to most of the
practical high-impedance surface structures.

Hopefully the results of this paper provide new insight into the
properties of mushroom structures as well as tools for the engineers
for efficient and accurate design work.


\end{document}